\documentstyle[12pt] {article}
 
\begin{document}
 \title{ {\small { \bf THE CP ASYMMETRY IN $b\rightarrow s{\it{l^{+}l^{-}}}$
 DECAY}}}
 \author{ {\small T.M.ALIEV,\,\,D.A.DEMIR,\,\,E.ILTAN,\,\,N.K.PAK}
 \\ {\it {\small Physics Department, Middle East Technical University}}
  \\ {\it {\small Ankara,Turkey}}}
 \begin{titlepage}
 \maketitle
 \thispagestyle{empty}
 \begin{abstract}
 \baselineskip  .7cm
 Using the experimental upper bound on the neutron EDM and experimental
 result on $b\rightarrow s\gamma$  branching ratio we have calculated CP
asymmetry and
 $\Gamma^{2HDM}(b\rightarrow s{\it{l^{+}l^{-}}})/\Gamma^{SM}(b\rightarrow
s{\it{l^{+}l^{-}}})$.
 It is shown that in the invariant dilepton mass $q^{2}$ region
 $(m_{\psi'}^{2}+0.2\; GeV^{2})<\;q^{2}<\;m^{2}_{b}$ the CP asymmetry
 is maximal and quite detectable.

 PACS numbers: 12.6.Fr, 11.30.Er, 11.30.Fs
 \end{abstract}
 \end{titlepage}
 \baselineskip  .7cm
  \newpage
 \section{Introduction}
The experimental discovery of the inclusive and exclusive decays
$B\rightarrow X_{s}\gamma$ and $B\rightarrow K^{*}\gamma$  by the
CLEO collaboration [1,2] has triggered a lot of theoretical and the
experimental
activity in the field of rare decays of B- mesons. These decays are interesting
for checking the predictions of SM at one-loop level, for determining the CKM
matrix
elements, and for looking for the "new physics" beyond the SM. From the
experimental
point of view another promising decay in this direction is the semileptonic
decay $b\rightarrow X_{s}{\it{l^{+}l^{-}}}$, because  this decay is easier to
measure
provided that we are given a good electromagnetic detector and a large number
of B hadrons.
Theoretically this decay has been the subject of many works in the
framework of the SM [3,4,5,6] and its extensions, particulary in Two Higgs
Doublet
Model (2HDM).

$b\rightarrow s{\it{l^{+}l^{-}}}$ decay is an FCNC process
which appears only at the one- loop level of pertubation theory. The basic
thing about this decay is that the penguin diagrams provide the two key
ingredients
needed for partial rate asymmetries. Being a loop diagram, it involves all
three generations, each generation contributing with different elements of
the CKM matrix. At the same time the loop effects that involve on- shell
particle rescatterings provide the necessary absorbtive parts.

It is well known that in 2HDM,  $b\rightarrow s{\it{l^{+}l^{-}}}$ decay
receives
significant contributions from the charged Higgs ($H^{\pm}$) exchange [7].
Another interesting pecularity of 2HDM is the appearence of new sources of
CP violation [8] in addition to the one in SM. An interesting version of 2HDM,
so called the most general 2HDM, which was proposed in [9], has a new source
of CP violation, arising from the relative phase between the vacuum
expectation values of two Higgs scalars.

In this work we shall work out $b\rightarrow s{\it{l^{+}l^{-}}}$ decay. In
particular we shall determine the CP asymmetry $A$ and the ratio
$r=\Gamma^{2HDM}(b\rightarrow s{\it{l^{+}l^{-}}})/\Gamma^{SM}(b\rightarrow
s{\it{l^{+}l^{-}}})$
as functions of the charged Higgs mass.

In the calculation of the CP asymmetry we shall consider both the SM and
2HDM contributions simultaneously. In determining $r$ and $A$ we shall make
use of the experimental results on $BR(b\rightarrow s\gamma)$ [1,2], and the
neutron electric dilpole moment (EDM).

Section 2 is devoted to the derivation of basic theoretical results and Section
3 contains the numerical analysis of them.
\section{Formalism}
In the most general 2HDM [8,9] the couplings of $H^{\pm}$ with $t_{R}$ and
$b_{R}$ is characterised by the coefficients $\xi_{f}$ defined by
\begin{eqnarray}
\xi_{f} =\frac{sin\delta_{f}}{sin\beta cos\beta
sin\delta}e^{i\sigma_{f}(\delta-\delta_{f})}- cot\beta
\end{eqnarray}
where f= t or b, $\sigma_{f}$ = + for b and - for t, and
$\delta_{f}=h_{2}/h_{1}$ where $h_{2}$ and $h_{1}$ are the diagonal
elements of the matrices $\Gamma^{u}_{2}$ and $\Gamma^{u}_{1}$ respectively.
Here $\Gamma^{u}$ are the matrices in the flavour space, and determine the
Yukawa couplings (for more detail see [9]), and $\delta$ is the relative
phase between the vacuum expectations of the two Higgs scalars:
\begin{eqnarray}
<\phi^{0}_{1}>&=&\frac{v}{\sqrt{2}}cos\beta e^{i\delta}\nonumber\\
<\phi^{0}_{2}>&=&\frac{v}{\sqrt{2}}sin\beta
\end{eqnarray}
The most general 2HDM reduces to the well-known 2HDM's in the current
literature,
in certain limiting cases [9]. Namely, if
$\delta_{t}=\delta_{b}=0$, then $\xi_{t}=\xi_{b}=-cot\beta$ (Model I) and,
if $\delta_{b}=\delta, \delta_{t}=0$, then $\xi_{t}=-cot\beta,
\xi_{b}=tan\beta$
(Model II).

As mentioned above the penguin diagrams provide the necessary absorbtive parts
for the calculation of the CP asymmetry. In this decay the dilepton invariant
mass $q^{2}$ ranges from $4m^{2}_{l}$ to $m^{2}_{b}$; therefore, u and c loops
give rise to nonzero absorbtive parts which are described, at the point
$\mu=m_{b}$, by
\begin{eqnarray}
F=i4\sqrt{2}G_{F}\lambda_{u}\frac{\alpha}{4\pi} A_{9}
\bar{s}_{L}\gamma_{\mu}b_{L}{\it{l}}^{+}\gamma_{\mu}{\it{l}}^{-}
\end{eqnarray}
where $\lambda_{i}=V_{is}V^{*}_{ib}$ and the function $A_{9}$ is given by
\begin{eqnarray}
A_{9}=w_{u}[Q(m^{2}_{c}/q^{2})-Q(m^{2}_{u}/q^{2})]
\end{eqnarray}
where
\begin{eqnarray}
Q(x)=\frac{2\pi}{9}(2+4x)\sqrt{1-4x}\theta(1-4x)
\end{eqnarray}
and $w_{u}$, having the numerical value of 0.3864, comes from the RGE movement
of the Wilson coefficients from $\mu = M_{W}$ to $\mu = m_{b}$ point.

It is well- known that in the range ($4m_{l}^{2}, m^{2}_{b}$) one can create
real low lying charmonium states [10,11]. In this work we shall discard that
portion of total dilepton mass range including $J/\psi$ and $\psi'$ poles
and the region between them to avoid the addition of new hadronic
uncertainities to the decay amplitude. Thus we restrict ourselves to the
following kinematical regions [6]:
\begin{eqnarray}
Region\; I &:& 4m^{2}_{{\it{l}}}\leq q^{2}\leq (m^{2}_{\psi}-\tau)\nonumber\\
Region\;II&:& (m^{2}_{\psi'}+\tau)\leq q^{2} \leq m^{2}_{b}
\end{eqnarray}
where $\tau = 0.2 GeV^{2}$ is the cut- off parameter.

Taking into account the 2HDM contributions and absorbtive part described by
$F$ in (3), the amplitude for $b\rightarrow s{\it{l^{+}l^{-}}}$ can be written
as
 \begin{eqnarray}
 M_{b\rightarrow s{\it{l}}^{+}{\it{l}}^{-}} & = &
4\sqrt{2}G_{F}\frac{\alpha}{4\pi}\times\nonumber\\
                                            &   & \{
C^{eff}_{9}(\mu)\bar{s}_{L}\gamma_{\mu}b_{L}{\it{l}}^{+}\gamma_{\mu}{\it{l}}^{-} + \nonumber\\
                                            &   &
C_{10}(\mu)\bar{s}_{L}\gamma_{\mu}b_{L}{\it{l}}^{+}\gamma_{\mu}\gamma_{5}{\it{l}}^{-} + \\
                                            &   & \frac{q^{\nu}}{q^{2}}\times
C_{7}(\mu)\bar{s}\sigma_{\mu\nu}(m_{b}R
+m_{s}L)b{\it{l}}^{+}\gamma_{\mu}{\it{l}}^{-} \}\nonumber
 \end{eqnarray}
The Wilson coefficients appearing in (7) are given by
 \begin{eqnarray}
C_{7}(\mu)&=&\lambda_{t}[C^{SM}_{7}(\mu) + C^{2HDM}_{7}(\mu)]\nonumber\\
C^{eff}_{9}(\mu)&=&\lambda_{t}[C^{SM}_{9}(\mu) + C^{2HDM}_{9}(\mu)]
+i\lambda_{u}A_{9}\\
C_{10}(\mu)&=&\lambda_{t}[C^{SM}_{10}(\mu) + C^{2HDM}_{10}(\mu)]\nonumber
\end{eqnarray}
The explicit forms of $C^{SM}_{i}(\mu)$, (i=7,9,10) including leading
and next-to-leading order QCD corrections can be found in [3,12,13,14].
The 2HDM contributions, $C^{2HDM}_{i}(\mu)$, in the framework of the most
general 2HDM [9] are given by
\begin{eqnarray}
C^{2HDM}_{7}(\mu)&=&\mid \xi_{t}\mid ^{2} K^{tt}_{7} + (R_{tb}
+iI_{tb})K^{tb}_{7}\nonumber\\
C^{2HDM}_{9}(\mu)&=&\mid \xi_{t}\mid ^{2} K^{tt}_{9}\\
C^{2HDM}_{10}(\mu)&=&\mid \xi_{t}\mid ^{2} K^{tt}_{10}\nonumber
\end{eqnarray}
where $R_{tb}=Re[\xi_{t}\xi^{*}_{b}],\;\; I_{tb}=Im[\xi_{t}\xi^{*}_{b}]$ and
\begin{eqnarray}
K^{tb}_{7} &=& \eta^{16/23}[G(y)-\frac{8}{3}(1-\eta^{-2/23})E(y)]\nonumber\\
K^{tt}_{7} &=&
\frac{1}{6}\eta^{16/23}[A(y)+\frac{8}{3}(1-\eta^{-2/23})D(y)]\nonumber\\
K^{tt}_{9} &=& -\frac{-1+4s^{2}_{W}}{s^{2}_{W}}\frac{x}{2}B(y)+yF(y)\\
K^{tt}_{10}&=&-\frac{1}{s^{2}_{W}}\frac{x}{2}B(y)\nonumber
\end{eqnarray}
with $x=m^{2}_{t}/M^{2}_{W}$, $y=m^{2}_{t}/M^{2}_{H}$, $s^{2}_{W}= 0.2315$,
$\eta = \alpha_{s}(M_{W})/\alpha_{s}(m_{b})$  and the explicit expressions
for functions $ A, B, D, E, F, G $ can be found in [12].

As noted in [9], $\xi_{t}$ is expected to be of order of unity or less, if the
Yukawa couplings of the top quark is reasonable. We have shown that this
happens to hold also for the decay process under consideration.
Thus, without loosing generality, in what follows we set $\mid \xi_{t}
\mid^{2}$ = 0 (all
the conclusions remain in force for the case of $\mid \xi_{t} \mid^{2}$ =1  as
well).

Using (7), the differential decay rate for $b\rightarrow s{\it{l^{+}l^{-}}}$ is
obtained as
\begin{eqnarray}
 \frac{d\Gamma^{2HDM}}{ds}&=& \lambda_{0}(1-s)^{2}\{ 4(\frac{2}{s}+1)\mid
C_{7}(\mu)\mid^{2}
 +(1+2s)(\mid C^{eff}_{9}(\mu)\mid^{2} +\mid C_{10}(\mu)\mid^{2})\nonumber\\
 &+&12 Re[C_{7}(\mu)C^{eff}_{9}(\mu)]\}
\end{eqnarray}
where $s=q^{2}/m^{2}_{b}$, and $\lambda_{0} =
\frac{\alpha^{2}G^{2}_{F}}{768\pi^{5}}$.

After integrating (11) over $s$ we get
\begin{eqnarray}
\gamma &=& \gamma_{0} +4\rho I^{2} + 2I(6I_{9}+6a^{(1)}_{9}R_{tu})\nonumber\\
       &+& 4\rho R^{2} + 2R(6R_{9}+6a^{(1)}_{9}I_{tu}+4\rho C_{7}^{SM})\\
       &+& 12a_{9}^{(1)}C^{SM}_{7}I_{tu}+a^{(2)}_{9}f_{tu}+
       2(a_{r9}I_{tu}+a_{i9}R_{tu})\nonumber
\end{eqnarray}
where
\begin{eqnarray}
\gamma&=&\frac{\Gamma^{2HDM}}{\lambda_{0}\mid \lambda_{t}\mid ^{2}}\nonumber\\
\gamma_{0}&=&(\frac{\Gamma^{SM}}{\lambda_{0}\mid \lambda_{t}\mid ^{2}})\mid
_{A_{9}=0}\nonumber\\
I&=&I_{tb}K^{7}_{tb}\\
R&=&R_{tb}K^{7}_{tb}\nonumber\\
I_{tu}&=&\frac{Im[\lambda_{t}\lambda_{u}^{*}]}{\mid \lambda_{t}\mid
^{2}}\nonumber\\
R_{tu}&=&\frac{Re[\lambda_{t}\lambda_{u}^{*}]}{\mid \lambda_{t}\mid
^{2}}\nonumber\\
f_{tu}&=&\frac{\mid \lambda_{u}\mid ^{2}}{\mid \lambda_{t}\mid ^{2}}\nonumber
\end{eqnarray}
and the other parameters in (12) are defined by the following integrals:
\begin{eqnarray}
\rho &=&\int ds (1-s)^{2}(\frac{2}{s}+1)\nonumber\\
R_{9}&=& \int ds (1-s)^{2}Re(C^{SM}_{9})\nonumber\\
I_{9}&=& \int ds (1-s)^{2}Im(C^{SM}_{9})\nonumber\\
a^{(1)}_{9}&=& \int ds (1-s)^{2}A_{9}\\
a^{(2)}_{9}&=& \int ds (1-s)^{2}(1+2s)A^{2}_{9}\nonumber\\
a_{r9}&=& \int ds (1-s)^{2}(1+2s)Re(C^{SM}_{9})A_{9}\nonumber\\
a_{i9}&=& \int ds (1-s)^{2}(1+2s)Im(C^{SM}_{9})A_{9}\nonumber
\end{eqnarray}

For the CP conjugate process, the analog of (12) can be obtained by the
following replacements:
\begin{eqnarray}
\bar{\gamma}=\gamma(I\rightarrow -I;\;\;I_{tu}\rightarrow -I_{tu})
\end{eqnarray}

Now we introduce the  parameter  $r$ that measures the relative strength
of 2HDM and SM rates
\begin{eqnarray}
r=\frac{\gamma}{\gamma_{SM}}
\end{eqnarray}
where $\gamma_{SM}$ is obtained by setting $I=R=0$ in (12).

Next we define the CP asymmetry by
\begin{eqnarray}
A=\frac{\bar{\gamma}-\gamma}{\bar{\gamma}+\gamma}
\end{eqnarray}

Substituting the expressions for $\gamma$  and $\gamma_{SM}$ into (16)
we obtain a circle for fixed values of $r$:
\begin{eqnarray}
(R+R_{0})^{2} +(I+I_{0})^{2} = t(r-1)+R_{0}^{2} +I_{0}^{2}
\end{eqnarray}
where the parameters $R_{0}$ and $I_{0}$ are given by
\begin{eqnarray}
R_{0}&=&\frac{3}{2\rho}(R_{9}+\frac{2}{3}\rho C^{7}_{SM})+r_{0}\nonumber\\
I_{0}&=&\frac{3}{2\rho}(I_{9}+a^{(1)}_{9}R_{tu})
\end{eqnarray}
and the quantity $r_{0}=\frac{3}{2\rho}a^{(1)}_{9}I_{tu}$ is introduced for
later use.

On the other hand, insertion of (12) and (15) into (17)  yields another
circle
\begin{eqnarray}
(R+R'_{0})^{2}+(I+I'_{0})^{2}=-t+\epsilon (1-\frac{1}{A})+
R'^{2}_{0} + I'^{2}_{0}
\end{eqnarray}
where
\begin{eqnarray}
I'_{0}&=&\frac{I_{0}}{A}\nonumber\\
R'_{0}&=&\frac{3}{2\rho}(R_{9}+\frac{2}{3}\rho C^{SM}_{7})+\frac{r_{0}}{A}
\end{eqnarray}
The parameters  $\epsilon$ and $t$ in (19) and (20) are given by
\begin{eqnarray}
\epsilon&=&\frac{I_{tu}}{4\rho}(12a^{(1)}_{9} C^{SM}_{7}+ 2a_{r9})\nonumber\\
t&=& -\frac{(1-A_{s})}{A_{s}}\epsilon
\end{eqnarray}
where $A_{s}$ is the CP asymmetry in SM which is obtained from (17) by:
\begin{eqnarray}
A_{s}=A\mid_{I=R=0}
\end{eqnarray}

Up to this point, our analysis of $b\rightarrow s{\it{l^{+}l^{-}}}$ decay
parallels that of $b\rightarrow s\gamma$ in [9] except for the definition
of $A$. We shall, however, analyze the circles in (18) and (20) in a different
context by exploiting the relation between $I$ and neutron EDM, and
experimental results on $b\rightarrow s\gamma$ branching ratio [1,2].

First we obtain the expression for the CP asymmetry in (17) by subtracting
(20) from (18) and solving for $A$:
\begin{eqnarray}
A=\frac{1}{1-a}
\end{eqnarray}
where
\begin{eqnarray}
a=\frac{tr}{\epsilon +2II_{0}+2Rr_{0}}
\end{eqnarray}

Now we turn to the determination of $I$ with the use of the experimental
upper bound on neutron EDM. Weinberg has proposed a CP violating 6
dimensional gluonic operator [15]
\begin{eqnarray}
O_{6}\sim f_{abc}G^{\mu \rho}_{a}G^{\nu}_{b\rho}\tilde{G}_{c\mu\nu}
\end{eqnarray}
which has been shown to give very
large contribution to neutron EDM, $d_{n}$ by the neutral [15] or
charged [16] Higgs exchange. Weinberg, after relating
the hadronic matrix elements of $O_{6}$ to $d_{n}$, predicts the
value of $d_{n}$ on the basis of a Naive Dimensional Analysis (NDA). However
a detailed analysis by Bigi and Uraltsev [17] reports a different value for
$d_{n}$ which equals $\frac{1}{30}$ of that of Weinberg's. The big
difference between the results of these analyses is an indication of the
existence of hadronic uncertainities which are mainly introduced by the
matrix elements of $O_{6}$ between the nucleon states. In addition to
these theoretical uncertainities,  we have also problems with experimental
data (in that experiment yields only an upper bound on neutron EDM).
These can be summarized as
\begin{eqnarray}
d^{theor}_{n}&=&c_{theor}\times I_{tb}K(y)10^{-25}\,e\,cm\\
d^{actual}_{n}&=&c_{exp}\times d^{max}_{n}
\end{eqnarray}
where $c_{theor}$ and $c_{exp}$  are constants and $\mid c_{exp}\mid $ is known
to
be less than unity. Let us note that $c_{theor}$ is related to the theoretical
uncertainities and $c_{exp}$ to the experimental uncertainities. Experiment
yields $d^{max}_{n}=1.1\;10^{-25} e\,cm$ [18]. The function  $K(y)$ in (27)
is given by [16,17]:
\begin{eqnarray}
K(y)=\frac{y}{(y-1)^{3}}[3/2-2y+y^{2}/2+ln(y)]
\end{eqnarray}
The common point for the analyses in [16] and [17] is the presence
of the function $K(y)$ which is equal to $\frac{1}{3}$ as $y\rightarrow 1$.

Equating (27) to (28) and defining $\beta=1.1\frac{c_{exp}}{c_{theor}}$,
we obtain
\begin{eqnarray}
I=\beta f(y)
\end{eqnarray}
where
\begin{eqnarray}
f(y)=\frac{K^{7}_{tb}(y)}{K(y)}
\end{eqnarray}

Note that the constant $\beta$ in (30) includes now both theoretical and
experimental
undeterminicies. We shall not make any assumption concerning the value of
$\beta$; instead we are going to fix it through the use of the experimental
results on $b\rightarrow s\gamma$ branching ratio.

The $b\rightarrow s\gamma$ decay amplitude is given by
\begin{eqnarray}
M=\frac{4G_{F}}{\sqrt{2}}\frac{\alpha}{4\pi}C_{7}(\mu)\bar{s}(p')
\sigma_{\mu\nu}(m_{b}R+m_{s}L)b(p)F^{\mu\nu}
\end{eqnarray}
where $C_{7}(\mu)$ is defined in (8). Using the experimental result on
the braching ratio of $b\rightarrow s\gamma$ decay [1,2] we  get the
following circle
\begin{eqnarray}
(C^{SM}_{7}+R)^{2}+I^{2}=(C^{ex}_{7})^{2}
\end{eqnarray}
where $C^{ex}_{7}$ is the experimental value of $C_{7}(\mu)$
\begin{eqnarray}
0.22 \leq \mid C^{ex}_{7}\mid \leq 0.30
\end{eqnarray}

We shall determine the central values of $\beta$, $r$ and $A$  which are
defined in equations (20), (16) and (17) respectively. In doing this, we
will make  use of circles in equations (18), (20) and (33) together with
equation (30). Let us note that (30) is obtained by the use of the
experimental upper bound on neutron EDM [18], and (33) is constructed with the
use of the experimental data on $b\rightarrow s\gamma$ branching ratio [1].

Let us first determine $\beta$.
For this purpose we consider the circle in (33) in the limit of infinitely
large $M_{H}$ or equivalently $y\rightarrow 0$. As $y\rightarrow 0$,
$R\rightarrow 0$ and through (30), $I\rightarrow \beta f_{0}$, where
numerically $f_{0}=0.2706$. Then equation (33), which is valid for any
value of $M_{H}$, yields
\begin{eqnarray}
\beta=\pm\{\frac{(C^{ex}_{7})^{2}-(C^{SM}_{7})^{2}}{f_{0}^{2}}\}^{1/2}
\end{eqnarray}

With (35), $I$ in (30) has now become a completely known function of
$M_{H}$. Now we solve (33) for $R$, yielding
\begin{eqnarray}
R=-C^{SM}_{7}+\sqrt{(C^{ex}_{7})^{2}-I^{2}}
\end{eqnarray}
where the choice of plus sign is necessary to satisfy asymptotic condition
on $R$.

Using (36) for $R$, and (30) for $I$ we can solve equation (18) for $r$
\begin{eqnarray}
r=1+\frac{(R+R_{0})^{2}+(I+I_{0})^{2}-R_{0}^{2}-I_{0}^{2}}{t}
\end{eqnarray}
whose $M_{H}$ dependence shall be discussed in the next section.

Finally, taking $r$ from (37), $R$ from (36) and $I$ from (30)
we determine the CP asymmetry $A$ in (24) whose dependence
on $M_{H}$ shall also be studied in the next section.

\section{Numerical Analysis}
In the numerical analysis we shall use $m_{u}=10 MeV,\;\;m_{c}=1.5 GeV
,\;\;m_{b}=4.6 GeV$. For the top quark mass we rely on the $CDF$ data [19]
and for the $W$ mass we use $M_{W}=80.22 GeV$ [18].

In calculating $I_{tu}$ and $R_{tu}$ we use the parametrisation in [18],
and in doing this we take the mid values of the quantities.
For the phase $\delta_{13}$ of $CKM$ matrix in [18] we shall use the
the mid value of $cos\delta_{13}=0.47\pm 0.32$ given in [20] which icludes
a large uncertainity. A straightforward calculation shows that corresponding
to the uncertainity in $cos\delta_{13}$, $R_{tu}$  and $I_{tu}$ are
uncertain by $3.87\%$  and $23.75\%$ respectively. Thus, the standard model
asymmtery $A_{s}$ in (23) is uncertain by $23.75\%$, and we shall use
its central value in our calculations. This choice is justified by the
closeness
of $I_{tu}$ and $R_{tu}$ calculated in this way to that obtained by the
use of Wolfenstein parametrisation [21].

Fig. 1 shows the variation of $f(y)$ in (29) with $M_{H}$
for the lowest, central and the highest values of $m_{t}$
permitted by the $CDF$ data [19]. As we see from Fig. 1
dependence of $f(y)$ on $m_{t}$ is very weak; thus,
insensitivity of results to the variation of $y$ with $m_{t}$
is guaranteed. In what follows we shall use therefore the central value
of $CDF$ data $m_{t}=176 GeV$.

For $m_{t}=176 GeV$ we obtain $C^{SM}_{7}=-0.2686$. The $b\rightarrow s\gamma$
branching ratio has approximately $50\%$ error [1] which is tranferred into a
range of values that $C^{ex}_{7}$ may take, as described by (34).

With the use of above-mentioned data we calculate SM CP asymmetry in (23)
to be $A_{s}=0.0714\%$ in Reg. I, and $A_{s}=0.0223\%$ in Reg. II.

In the second column of Table 1 we give the values of $\beta$ as
$\mid C^{ex}_{7}\mid$ moves from its maximum value 0.30 towards
$\mid C^{SM}_{7}\mid=0.2686$. We see that $\mid \beta \mid $
decreases gradually with decreasing $\mid C^{ex}_{7}\mid$. Moreover,
it is seen that the maximum value that $\mid \beta \mid \approx 0.5$.

Regarding the present calculations in [16] and [17] as the possible candidates
for $c_{theor}$ in (27), we can make certain predictions for $c_{exp}$
in (28). A simple calculation yields $c_{theor}=9.9$ and $c_{theor}=0.33$
for Weinberg's NDA and Bigi-Uraltsev calculations respectively. In the
case of NDA, a solution for $c_{exp}$ exist only for $\mid\beta\mid <\sim 0.27$
at which $d^{actual}_{n}$ turns out to be very close to its experimental upper
bound. On the other hand, for Bigi-Uraltsev calculation, being a more detailed
analysis,
for all values of $\mid C^{ex}_{7}\mid $ ranging from $\mid C^{SM}_{7}\mid$ to
0.30 there
exists a solution for $c_{ex}$ with the help of which, through (28), one
determine the value $d^{actual}_{n}$. In the third column of Table 1 we give
the values of
$d^{actual}_{n}$  as $\mid C^{ex}_{7}\mid$ moves from its maximum value
0.30 towards $\mid C^{SM}_{7}\mid=0.2686$. We observe that for
$\mid C^{ex}_{7}\mid=0.3$ $\mid d^{actual}_{n} \mid$ reaches its maximum
value of $1.63\;10^{-26}$ which is one order of magnitude less than the
present experimental upper bound.

In our numerical analysis we use the range of values of $M_{H}$ from $44 GeV$
[18]
to $10 m_{t}$ [15]. In Fig. 2 and Fig.3 we show the variation of $r$ in (37)
with
$M_{H}$ in Regions $I$ and $II$ respectively. We observe that in both
figures $r$ is fairly high at low $M_{H}$ and lands rapidly to a lower
value after $M_{H}\sim 500 GeV$.

As we see from Fig.2, dependence of $r$ on the sign of $\beta$ in Region $I$
is very weak. Moreover, for $M_{H}>\sim 1 TeV$, $r$ attains the values $\sim
1.056$,
$\sim 1.0050$, $\sim 1.020$, and $\sim 1.016$ for $\beta=+0.4938,\;
-0.4938,\; 0.2922$, and $-0.2922$ respectively.

 From Fig.3 we observe that in Region $II$ dependence of $r$ on the sign of
$\beta$ is large. Specificially, we see that, for large $M_{H}$, $r$
becomes practically independent of $M_{H}$ and attains the values
$\sim 1.021,\; \sim 0.998,\; \sim 1.01$, and $\sim 0.9996$ corresponding to
$\beta=+0.4938,\;-0.4938,\; 0.2922$ and $-0.2922$ respectively.

In Fig.4 and Fig.5 we show the variation of $A$ in (24) with $M_{H}$ in
Regions $I$ and $II$ respectively. What we observe to be common between
them is the saturation of CP asymmetry $A$ to a certain value after
$M_{H}\sim 500 GeV$.

 From Fig.4 we observe that the $2HDM$ CP asymmetry $A$, practically for all
$M_{H}$, is of the same order as the SM CP asymmetry $A_{s}$.
Indeed, especially for large $M_{H}$, corresponding to the values of $\beta$,
$\beta=+0.4938,\;-0.4938,\; 0.2922$ and $-0.2922$, $A$ attains the
percentage values of $\sim -0.27,\; \sim 0.40,\; \sim -0.14$,
and $\sim 0.28$.

In Fig. 5 we observe that asymmetry $A$, as compared to the previous figure,
is completey different in that it is positive and takes higher values
for all values of $M_{H}$. Actually, we see that for small $M_{H}$,
$2HDM$ CP asymmetry is larger than the SM CP asymmetry by approximately
three orders of magnitude. For large $M_{H}$, however, $A$ gets values
which are larger than SM asymmetry by two orders of magnitue.
Indeed, for large $M_{H}$, corresponding to the values of $\beta$,
$\beta =+0.4938,\;-0.4938,\; 0.2922$ and $-0.2922$, $A$ gets the
following percentage values $\sim 1.1,\; \sim 3.25,\; \sim 0.2$, and $\sim 1.5$

The last point to be noted about the Figs. 2-5 is that negative $\beta$ gives
rise to larger $r$ and $A$ than positive $\beta$ does.

To decern a CP asymmetry $A$ at the $\sigma$ significance level with only
statistical
errors, the number of $B$ hadrons $N_{B}$ needed to demonstrate the asymmetry
is given
by[22]
\begin{eqnarray}
N_{B}\approx \frac{\sigma^{2}}{BR\times A^{2}}
\end{eqnarray}
Now denoting the number of $B$ hadrons to observe $A_{s}$, $A$ in $I$ and
$A$ in $II$ by $N_{B}^{s}$, $N_{B}^{I}$ and $N_{B}^{II}$ respectively, we
get, using the values of $r$ and $A$ we have obtained already,
the following ratios
\begin{eqnarray}
\frac{N_{B}^{I}}{N_{B}^{s}}&\approx& 1\nonumber\\
\frac{N_{B}^{II}}{N_{B}^{s}}&\approx& 10^{-4}
\end{eqnarray}
which clearly  prove that  Region $II$ is more suitable for
experimental investigations on $A$.

In conclusion we have determined the 2HDM CP asymmetry $A$, ratio of 2HDM decay
rate
to SM decay rate $r$ and actual value of neutron EDM. In doing these we have
utilized the experimental results on $b\rightarrow s\gamma$ branching ratio,
and on the upper bound of neutron EDM. Both $r$ and $A$  relax to constant
values after $M_{H}\sim 500 GeV$. This saturation property of quantities
shows that if charged Higgs mass happens to be large ($\sim 1 TeV$) then
the most general $2HDM$ merely shifts the SM values of $r$ and $A$ to some
other value which may be important for establishing 2HDM. Boldly speaking,
in the high dilepton mass region (Region $II$) $r$ is closer to unity and
asymmetry is very large as compared to those in low dilepton mass region
(Region $I$). Thus on the basis of the order of magnitude analysis carried out
for $N_{B}$, we conclude that the high dilepton mass region is important and
appropriate for experimental check of the quantities under concern. Region $II$
[6]
is accessible to the $B$ experiments which will be carried out with hadron
beams in CDF, HERA and LHC.

     \newpage
{ \bf Figure Captions }
 \begin{description}
\item[{\bf Figure 1}:] The $M_{H}$ dependence of $f(y)$ for $m_{t}=194 GeV$
(with circles ), $m_{t}=176 GeV$ (bare solid curve) and $m_{t}=158 GeV$
(with squares).
\item[{\bf Figure 2}:] The $M_{H}$ dependence of $r$ in Region $I$.
Here labes 1, 2, 3 and 4 correspond to $\beta=0.4938,\;-0.4938,\;0.2922$
and $-0.2922$ respectively.
\item[{\bf Figure 3}:] The same as in Fig. 2 but for Region $II$.
\item[{\bf Figure 4}:] The $M_{H}$ dependence of $A$ in Region $I$.
Labels have the same meaning as in Fig.1. Here the unlabled solid line
shows the SM asymmetry.
\item[{\bf Figure 5}:] The same as in Fig. 4 but for Region $II$.
\end{description}
\end{document}